\documentclass[preprints,review,accept,moreauthors,pdftex]{Definitions/mdpi}

\allowdisplaybreaks

\usepackage{mathrsfs} 
\usepackage[mathscr]{euscript}
\usepackage{amssymb,amsbsy,amsmath,amsfonts}
\usepackage{slashed}
\usepackage{nicefrac}

\def\beq{\begin{equation}}
\def\eeq{\end{equation}}
\def\bea{\begin{eqnarray}}
\def\eea{\end{eqnarray}}

\def\eqlab#1{\label{eq:#1}}

\def\eref#1{(\ref{eq:#1})}
\def\Eqref#1{Eq.~(\ref{eq:#1})}

\def\boxfrac#1#2{\mbox{$\frac{#1}{#2}$}}
\def\al{\alpha}
\def\be{\beta}
\def\ga{\gamma} 
\def\de{\delta} \def\De{\Delta}
\def\veps{\varepsilon}  \def\eps{\epsilon}

\def\si{\sigma} \def\Si{{\Sigma}}
  
\def\w{\omega}

\def\scA{\mathscr{A}}
\def\scO{\mathscr{O}}

\def\nn{\nonumber}
\def\dd{\mathrm{d}}

\def\ol#1{\overline{#1}}
\def\half{\mbox{\small{$\frac{1}{2}\;$}}}

\firstpage{1} 
\pubvolume{xx}
\issuenum{1}
\articlenumber{5}
\pubyear{2019}
\copyrightyear{2019}
\history{}

\pdfoutput=1

\Title{Nucleon Polarizabilities and Compton Scattering as Playground for Chiral Perturbation Theory}

\Author{Franziska Hagelstein $^{1}$\orcidA{}}

\AuthorNames{Franziska Hagelstein}

\address{%
$^{1}$ \quad Albert Einstein Center for Fundamental Physics, Institute for Theoretical Physics, University of Bern, Sidlerstrasse 5, CH-3012 Bern, Switzerland; hagelstein@itp.unibe.ch}

\abstract{I give a summary of recent results on nucleon polarizabilities, with emphasis on chiral perturbation theory. The predictive calculations of Compton scattering off the nucleon are compared to recent empirical determinations and lattice QCD calculations of the  polarizabilities, thereby testing chiral perturbation theory in the single-baryon sector. }


\begin{document}

\section{Introduction}

The name Chiral Perturbation Theory ($\chi$PT) was first introduced in the seminal works of \citet{Pagels:1974se}, who used
it to describe a systematic expansion in the pion mass
$m_\pi$, which is small compared to other hadronic scales.
Some years later, in 1979, Weinberg \cite{Weinberg:1978kz} made an enlightening  proposal for effective-field theories (EFT) and
the $\chi$PT acquired its present meaning by Gasser and Leutwyler  \cite{Gasser:1983yg,Gasser:1987rb} in this, more powerful, connotation. Since then, $\chi$PT stands for a low-energy EFT of the strong sector of the Standard Model. Written in terms of hadronic degrees of freedom, rather than quarks and gluons, it offers
an efficient way of calculating low-energy hadronic physics. 
Many calculations can be done analytically in a systematic perturbative expansion, in contrast 
to the \textit{ab initio} calculations, viz., lattice QCD, Dyson-Schwinger equations, and other non-perturbative
calculations in terms of quark and gluon fields.  

\begin{figure}[t]
\centering
\includegraphics[width=5 cm]{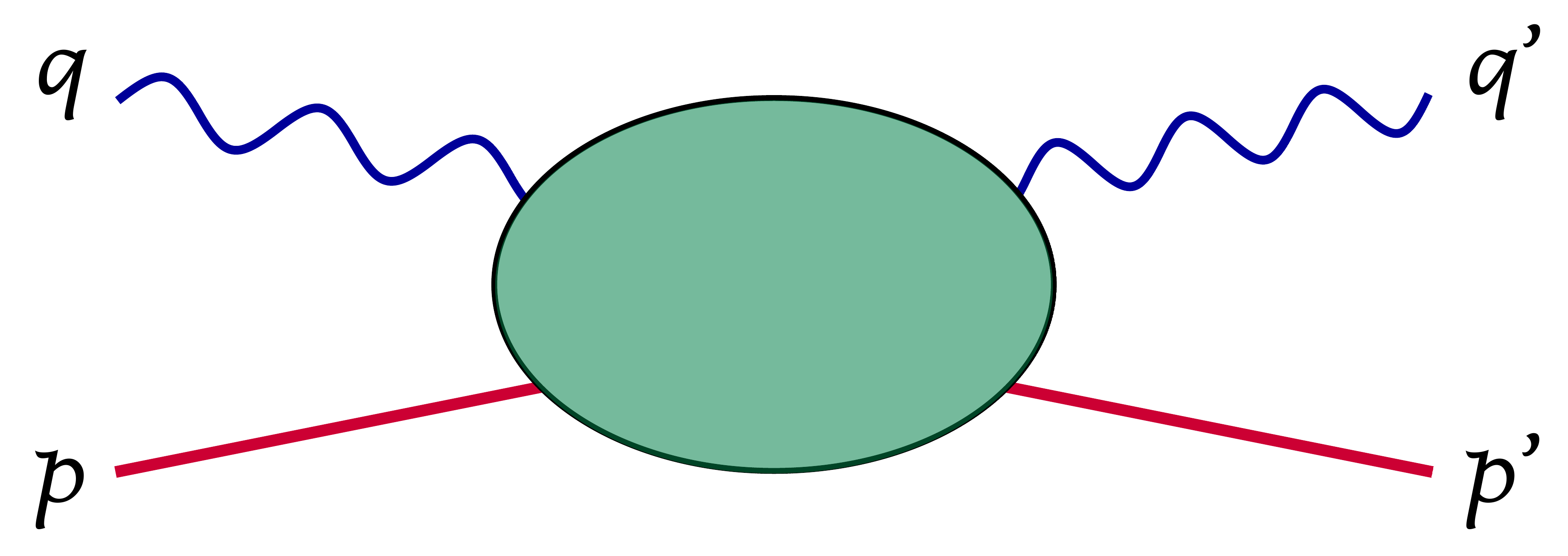}
\caption{CS off the nucleon in general kinematics: $\gamma^*(q) N(p) \rightarrow \gamma^*(q') N(p')$. \label{CSfigure}}
\end{figure} 

However, as in any EFT framework, the convergence and the predictive power of $\chi$PT calculations are often of concern.
After all, the expansion in energy and momenta is not
as clear-cut as usual expansions in a small coupling constant. And, each new order brings more and more free parameters --- the low-energy constants (LECs). This is why the
cases where $\chi$PT provides true predictions are very valuable. One such case, considered here, is the process of Compton scattering (CS) off the nucleon, see Figure \ref{CSfigure}. It allows one to study the low-energy properties of the nucleon \cite{Bernard:1991rq,Bernard:1991ru}.

The nucleon is characterized by a number
of different polarizabilities, the most important of which 
are the electric and magnetic dipole polarizabilities $\alpha_{E1}$ and $\beta_{M1}$. These quantities describe the size of the electric and magnetic dipole moments induced by an external electric $\vec{E}$ or magnetic $\vec{H}$ field: \begin{subequations}
\bea
\vec{d}_\mathrm{ind.} &=& 4\pi \alpha_{E1} \vec{E},\\
\vec{\mu}_\mathrm{ind.} &=& 4\pi \beta_{M1} \vec{H}.
\eea
\end{subequations}
In loosely bound systems, such as atoms and molecules, these
polarizabilities are roughly given by the volume of the system. The nucleon is apparently a much more rigid object --- its polarizabilities are orders of magnitude smaller than its volume ($\sim 1$ fm$^3$). The most accurate evidence of that comes from the Baldin sum rule (sometimes referred to as the Baldin--Lapidus sum rule) \cite{Baldin:1960,Lapidus1962}.
It is a very general relation based on the principles of causality, unitarity and crossing symmetry akin to the Kramers--Kronig relation (see, e.g., Ref.~\cite{Pascalutsa:2018ced} for a pedagogical review). The Baldin 
sum rule expresses the sum of dipole polarizabilities in terms of an integral of the total photoabsorption cross section $\si_T$:
\beq
\al_{E1}+\beta_{M1}=\frac{1}{2 \pi^2} \int_{\nu_0}^\infty \! \dd\nu\, \frac{\sigma_T (\nu)}{\nu^2}.\eqlab{BaldinSumRule}
\eeq
 Empirical evaluations \cite{Damashek:1969xj,Schroder:1977sn,Babusci:1997ij,Levchuk:1999zy,Olm01,Gryniuk:2015eza}, based on experimental cross sections of total photoabsorption on the nucleon, yield the most accurate information on proton and neutron dipole polarizabilities we presently have:
\begin{subequations}
\label{BaldinSRresults}
\bea
\al_{E1p}+\be_{M1p} &=& 14.0(2)\times 10^{-4}\,\text{fm}^3\qquad \text{\cite{Gryniuk:2015eza}},\\ 
\al_{E1n}+\be_{M1n} &=& 15.2(4)\times 10^{-4}\,\text{fm}^3\qquad \text{\cite{Levchuk:1999zy}}.
\eea
\end{subequations}

To disentangle $\alpha_{E1}$ and $\beta_{M1}$, one measures the angular distribution of low-energy CS. For example, the low-energy expansion of the unpolarized CS cross section is given by [to $\mathcal{O}(\nu^2)$]:
\beq
\frac{\dd \sigma}{\dd\Omega_L}-\frac{\dd \sigma^\text{Born}}{\dd\Omega_L}=-\nu \nu' \left(\frac{\nu'}{\nu}\right)^2 \frac{2\pi \al}{M_N}\left[\left(\al_{E1}+\be_{M1}\right)(1+\cos \theta_L)^2+\left(\al_{E1}-\be_{M1}\right)(1-\cos \theta_L)^2\right],
\eeq
where $\theta_L$ is the scattering angle, $\dd\Omega_L=2\pi\, \dd\! \cos\theta_L$, and $\nu(\nu')$ is the energy of the incoming (scattered) photon, all in the lab frame. Here, in addition to the sum of dipole polarizabilities appearing in forward kinematics, one can measure their difference. Another interesting observable is the beam asymmetry $\Sigma_3$ defined in Eq.\ (\ref{Eq:sigma_3}), which also provides access to $\beta_{M1}$ independent of $\alpha_{E1}$ at $\mathcal{O}(\nu^2)$, cf.\ Eq.~(\ref{Eq:krupina}).

In reality, the CS data are taken at finite energies (typically around 100 MeV), rather
than at infinitesimal energies required for a strict validation of the above low-energy expansion.  For a model-independent empirical extraction of polarizabilities from the RCS data
it is, therefore, important to have a systematic theoretical framework such as $\chi$PT or a partial-wave analysis (PWA).

\begin{figure}[t]
\centering
\includegraphics[width=0.9\columnwidth]{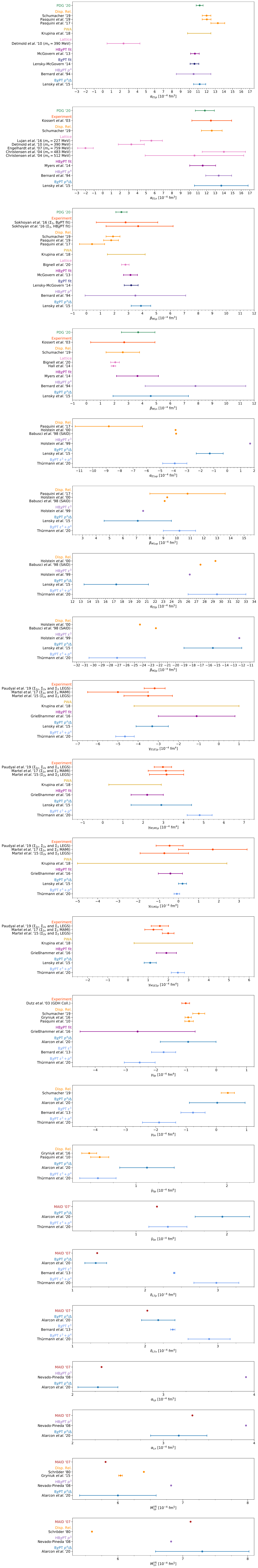}
\caption{Summary for the electric dipole polarizability of the proton $\al_{E1p}$ (upper panel) and neutron $\al_{E1n}$ (lower panel). Theoretical predictions from chiral EFT and lattice QCD are compared with extractions based on CS data. Note that the lattice QCD calculations are done at unphysical pion masses. For the proton one observes a small tension between the dispersive approaches to CS and the B$\chi$PT results.\label{alphaE1Pol}}
\end{figure}

\begin{figure}[t]
\centering
\includegraphics[width=0.9\columnwidth]{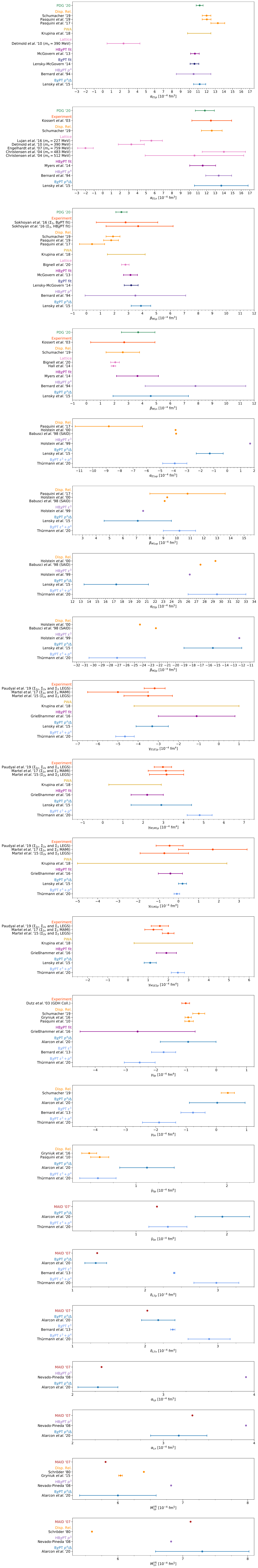}
\caption{Summary for the magnetic dipole polarizability of the proton $\be_{M1p}$ (upper panel) and neutron $\be_{M1n}$ (lower panel). Theoretical predictions from chiral EFT and lattice QCD are compared with extractions based on CS data. Note that the lattice QCD results are extrapolated to the physical pion mass. For the proton one observes a small tension between the dispersive approaches to CS and the B$\chi$PT results.  \label{betaM1Pol}}
\end{figure}

There are other interesting polarizabilities, called the spin polarizabilities. These are more difficult to visualise in a classical picture, but they certainly characterize the spin
structure of the nucleon.
$\chi$PT provides robust predictions for the different nucleon polarizabilities at leading and next-to-leading order. Given
the accurate empirical knowledge 
of the nucleon polarizabilities from dispersive sum rules and CS experiments,  they become an important benchmark for $\chi$PT in the single-baryon sector. But not just for $\chi$PT --- the lattice QCD studies of nucleon polarizabilities are also closing in on the physical pion mass, see Figures~\ref{alphaE1Pol} and \ref{betaM1Pol}.

It is worth mentioning that $\chi$PT can be used for calculating the proton-structure corrections to the muonic-hydrogen spectrum. These corrections are not only relevant in the context of the proton-radius puzzle \cite{Pohl:2010zza,Antognini:1900ns}, but also for the planned measurements of the muonic-hydrogen ground-state hyperfine splitting \cite{Pohl:2016xsr,Bakalov:2015xya,Kanda:2018oay}. 
The $\chi$PT is thusfar the only theoretical framework which can reliably
compute the polarizability effects in CS observables and, at the same time, in atomic spectroscopy.
In this way, a calculation which is validated on experimental data of CS and photoabsorption (through sum rules) can be used to predict the effects in muonic hydrogen
\cite{Alarcon:2013cba,Hagelstein:2015lph,Hagelstein:2017cbl}.

This mini-review is by no means comprehensive. A more proper review can be found in  Ref.~\cite{Hagelstein:2015egb}, whereas here I primarily provide an update on the nucleon polarizabilities. For the reader interested in the update only, I recommend to skip to Section \ref{results} where a description of all summary plots is given. A recent theoretical discussion of nucleon polarizabilities in $\chi$PT and beyond can be found in Ref.~\cite{Lensky:2019pye}. Other commendable reviews include: \citet{Guichon:1998xv} or \citet{Fonvieille:2019eyf} (VCS and generalized polarizabilities), \citet{Drechsel:2002ar} or \citet{Pasquini:2018wbl} (dispersion relations for CS), \citet{Pascalutsa:2006up} ($\Delta(1232)$ resonance), \citet{Phillips:2009af} (neutron polarizabilities), \citet{Griesshammer:2012we} ($\chi$EFT and RCS experiments), \citet{Holstein:2013kia} (pion, kaon, nucleon polarizabilities), \citet{Geng:2013xn} (B$\chi$PT), \citet{Pascalutsa:2018ced} (dispersion relations), \citet{Deur:2018roz} (nucleon spin structure). A textbook introduction to $\chi$PT can be found in Ref.~\cite{Scherer2012}.

The paper is organized as follows. In Sections \ref{BChPTintro} and \ref{CSintro}, I briefly describe the $\chi$PT framework and the CS formalism. In Section \ref{results}, I summarize recent $\chi$PT results for the nucleon polarizabilities and compare to empirical and lattice QCD evaluations.

\section{Baryon Chiral Perturbation Theory} \label{BChPTintro}

The low-energy processes involving a nucleon, such as $\pi N$ scattering or CS off the nucleon, can be described by SU(2) baryon chiral perturbation theory (B$\chi$PT), which is the manifestly Lorentz-invariant variant of $\chi$PT in the single-baryon sector \cite{Gasser:1987rb,Gegelia:1999gf,Fuchs:2003qc}. To introduce it, I will start in Section \ref{NuclPC} with the basic EFT including only pions and nucleons. Then, in Section \ref{DeltaPC}, I will discuss different ways (counting schemes) for incorporation of the lowest nucleon excitation --- the $\Delta(1232)$ resonance --- into the $\chi$PT framework. In Section \ref{LECsec}, I will show how the LECs can be fit to experimental data and discuss the predictive power of $\chi$PT for CS. In Section \ref{secHB}, I introduce the heavy-baryon chiral perturbation theory (HB$\chi$PT) and point out how its predictions differ from B$\chi$PT for certain polarizabilities. For more details on B$\chi$PT for CS, I refer to the following series of calculations: RCS \cite{Lensky:2008re,Lensky:2009uv,Lensky:2015awa}, VCS \cite{Lensky:2016nui} and forward VVCS \cite{Lensky:2014dda,Alarcon:2020wjg, Alarcon:2020icz}.

\subsection{B$\chi$PT with Pions and Nucleons}
\label{NuclPC}
 Consider the basic version of SU(2) B$\chi$PT including only pion and nucleon fields \cite{Gasser:1987rb}: scalar iso-vector $\pi^a(x)$ and spinor iso-doublet $\mathcal{N}(x)$. Expanding the EFT Lagrangian \cite{Gasser:1987rb}  to leading orders in pion derivatives,  mass and fields, one finds (see, e.g., Ref.~\cite{Ledwig:2011cx}):
\begin{subequations}
\eqlab{PionLagrangians}
\bea
\mathcal{L}^{(1)}_\mathcal{N}&=&\ol{\mathcal{N}} \big(\slashed{D}-M_N\big)\mathcal{N}-\frac{g_A}{2f_\pi}\ol{\mathcal{N}}\, \tau^a \left(\slashed{D}^{ab} \pi^b\right)\gamma_5\, \mathcal{N},\eqlab{piNN}\\
\mathcal{L}^{(2)}_\pi&=&\frac{1}{2} \Big(D_\mu^{ab} \pi^b\Big)\Big(D^\mu_{ac} \pi^c\Big)-\frac{1}{2}m_\pi^2 \pi_a\pi^a,\eqlab{pionLag}
\eea
\end{subequations}
with the covariant derivatives:
\begin{subequations}
\bea
D_\mu^{ab} \pi^b&=&\delta^{ab} \partial_\mu \pi^b+ieQ_\pi^{ab}A_\mu\pi^b,\\
D_\mu \mathcal{N}&=&\partial_\mu \mathcal{N}+ieQ_N A_\mu\mathcal{N}+\frac{i}{4f_\pi^2}\,\varepsilon^{abc}\tau^a \pi^b (\partial_\mu \pi^c),
\eea
\end{subequations}
the photon vector field $A_\mu(x)$,
and the charges:
\begin{subequations}
\bea
Q_\pi^{ab}&=&-i \varepsilon^{ab3},\\
Q_N&=&\half \!(1+\tau^3).
\eea
\end{subequations}
Here, $\tau^a$ are the Pauli matrices, $\ga_5 = i \ga^0\ga^1\ga^2\ga^3$ are the Dirac matrices, $\varepsilon^{ijk}$ is the Levi-Cevita symbol, and all other parameters are introduced in Table \ref{LECtable}. 

The key ingredient for the development of $\chi$PT as a low-energy EFT of QCD was the 
observation that the pion couplings are proportional to their four-momenta \cite{Weinberg:1978kz,Gasser:1983yg,Gasser:1987rb}. Therefore, at
low momenta the couplings are weak and a perturbative expansion is possible. This chiral expansion is done in powers of pion momentum and mass, commonly denoted as $p$, over the scale of spontaneous chiral symmetry breaking, $\Lambda_{\chi\mathrm{SB}}\sim 4 \pi f_\pi \approx 1$ GeV.  Therefore, one expects that $\chi$PT provides a systematic description of the strong interaction at energies well below $1$ GeV. Considering only pion and nucleon fields, the chiral order $\mathcal{O}(p^{n})$ of a Feynman diagram with $L$ loops, $N_\pi$ ($N_N$) pion (nucleon) propagators, and $V_k$ vertices from $k$-th order Lagrangians [e.g., $k=1$: ${\gamma NN}$ interaction from \Eqref{piNN}, $k=2$: ${\gamma\pi\pi}$ interaction from \Eqref{pionLag}] is defined as \cite{Gasser:1987rb}:
\beq
n=4L-2N_\pi-N_N+\sum_k k\, V_k.
\eeq

In the case of CS, the low-energy scale $p$ also includes the photon energy $\nu$ and virtuality $Q$, which therefore should be much smaller than $1$ GeV. 
However, the presence of bound states or low-lying resonances may lead to a breakdown of this perturbative expansion.   For example, in $\pi \pi$ scattering the limiting scale of the perturbative expansion is set by the $\sigma(600)$ and $\rho(775)$ mesons  \cite{Colangelo:2001df,Caprini:2005an}. In the single-nucleon sector, the breakdown
scale is set by the excitation energy of the first nucleon resonance, the $\Delta(1232)$ isobar.  
That is unless the $\Delta(1232)$ is included explicitly in the effective Lagrangian.

\begin{table}[h]
\caption{LECs and other parameters and the orders at which they appear in the chiral expansion when employing the low-energy $\delta$-expansion counting scheme. \label{LECtable}}
\centering
\begin{tabular}{ccccc}
\toprule
\textbf{Order in} &	\multicolumn{2}{c}{}&	& \\
\textbf{chiral expansion} &\multicolumn{2}{c}{\multirow{-2}{*}{\textbf{$\chi$PT parameters}}}	& \multirow{-2}{*}{\textbf{Values}}	&\multirow{-2}{*}{ \textbf{Sources}}\\
\midrule
 &fine-structure constant& $\al=\nicefrac{e^2}{4\pi}$ & $\simeq 1/137.04$ &\\
\multirow{-2}{*}{$\mathcal{O}(p^2)$ }&nucleon mass& $M_N$ & $938.27$ MeV &\\
\hline
&nucleon axial charge&$g_A$&$1.27$&neutron decay $n\rightarrow p\,e^-\, \bar\nu_e$ \cite{Olive:2016xmw}\\
 &pion decay constant&$f_\pi$&$92.21$ MeV&pion decay $\pi^+\rightarrow \mu^+ \nu_\mu$ \cite{Olive:2016xmw}\\
 \multirow{-3}{*}{$\mathcal{O}(p^3)$} &  pion mass&$m_\pi$ & $139.57$ MeV &\\
 \hline
 &&&&$P_{33}$ partial wave in $\pi N$ scattering\\
&\multirow{-2}{*}{$\mathcal{N}$-to-$\Delta$ axial coupling}&\multirow{-2}{*}{$h_A$}&\multirow{-2}{*}{$2.85$}&and $\Delta(1232)$ decay width \cite{Pascalutsa:2006up,Pascalutsa:2004je,Pascalutsa:2005nd} \\
 &$\Delta(1232)$ mass& $M_\Delta$ & $1232$ MeV &\\
&magnetic (M1) coupling &$g_M$&$2.97$ &\\
& electric (E2) coupling &$g_E$&$-1.0$&\\
\multirow{-6}{*}{$\mathcal{O}(p^4/\varDelta)$} &Coulomb (C2) coupling &$g_\mathrm{C}$&$-2.6$&\multirow{-3}{*}{\shortstack{pion electroproduction\\ $e^- N \rightarrow e^- N \pi$ \cite{Pascalutsa:2005vq}}}\\
\bottomrule
\end{tabular}
\end{table}
\subsection{Inclusion of the $\Delta(1232)$ and Power Counting}\label{DeltaPC}

The $\Delta(1232)$ resonance as the lightest nucleon excitation has an excitation energy
\beq
\varDelta = M_\Delta - M_N \simeq 293\, \mathrm{MeV},
\eeq
which is of the same order of magnitude as the pion mass. In the following, it will be included as an explicit degree of freedom: vector-spinor iso-quartet $\Delta_\mu(x)$. The relevant Lagrangians read~\cite{Ledwig:2011cx,Pascalutsa:2005ts,Pascalutsa:2005vq}:\footnote{At higher orders one also needs
\beq
\mathcal{L}^{(1)}_\Delta=\ol{\Delta}_\mu \left(i \gamma^{\mu \nu \lambda}D_\lambda -M_\Delta \gamma^{\mu \nu}\right) \Delta_\nu+\frac{H_A}{2f_\pi M_\Delta}\,\varepsilon^{\mu \nu \al \lambda}\,\ol \Delta_\mu \mathscr{T}^{\,a} \left(D_\al \Delta_\nu\right)D_\lambda^{ab}\pi^b,
\eeq
with $H_A=\nicefrac{9}{5}\,g_A$ the axial charge of the $\Delta(1232)$.}
\begin{subequations}
\eqlab{DeltaLagrangians}
\bea
\mathcal{L}^{(1)}_{\pi\Delta \mathcal{N}}&=&\frac{i h_A}{2f_\pi M_\Delta} \ol{\mathcal{N}} \,T^a\gamma^{\mu \nu \lambda}\left(D_\mu \Delta_\nu\right)\left(D_\lambda^{ab}\pi^b\right)+\text{h.c.},\\
\eqlab{nmGammaNDeltaLag}
 \mathcal{L}^{(2) \,\mathrm{non-minimal}}_{\gamma N \Delta}&=&\frac{3e}{2M_N(M_N+M_\Delta)}\Bigg[\ol{\mathcal{N}}T_3\Big\{i g_M (\partial_\mu \Delta_\nu) \tilde{F}^{\mu \nu}-g_E  \gamma_5 (\partial_\mu \Delta_\nu) F^{\mu \nu}\nn\\
 &&+i \frac{g_\mathrm{C}}{M_\Delta}\gamma_5 \gamma^\al (\partial_\al \Delta_\nu-\partial_\nu \Delta_\al)\partial_\mu F^{\mu \nu}\Big\}+\text{h.c.}\Bigg],\eqlab{LagrangianGND}
\eea
\end{subequations}
with the
covariant derivative:
\beq
D_\mu \Delta_\nu= \partial_\mu \Delta_\nu+ie Q_\Delta A_\mu \Delta_\nu+\frac{i}{2f_\pi^2} \,\varepsilon^{abc}\, \mathscr{T}^a\pi^b(\partial_\mu \pi^c),
\eeq
and the
charge:
\beq
Q_\Delta=\half \!(1+3 \mathscr{T}^3).
\eeq
Here, h.c.\ stands for the hermitian conjugate, $\gamma^{\mu \nu}=-\frac{i}{2}\epsilon^{\mu \nu \al \be}\ga_\al \ga_\be \ga^5$ and $\gamma^{\mu \nu \al}=-i\epsilon^{\mu \nu \al \be} \ga_\be \ga^5$ are Dirac matrices with $\epsilon_{0123}=1$, $F^{\mu\nu}=\partial^\mu A^\nu-\partial^\nu A^\mu$ is the electromagnetic field strength tensor, $\tilde F^{\mu\nu}=\epsilon^{\mu\nu\rho\lambda}\partial_\rho A_\lambda$ is its dual, and $T^a$ ($\mathscr{T}^a$) are the isospin $1/2$ ($3/2$) to $3/2$ transition matrices. The latter commute with the Dirac matrices. The superscripts of the Lagrangians in Eqs.~\eref{PionLagrangians} and \eref{DeltaLagrangians} denote their order as reflected by the number of comprised small quantities: pion mass, momentum and factors of $e$. Inclusion of the $\Delta(1232)$ introduces the excitation energy $\varDelta$ as another small scale, which has to be considered when defining a power-counting for the perturbative $\chi$PT expansion.

\begin{figure}[t]
\centering
\hspace{-2cm}\includegraphics[width=0.9\columnwidth]{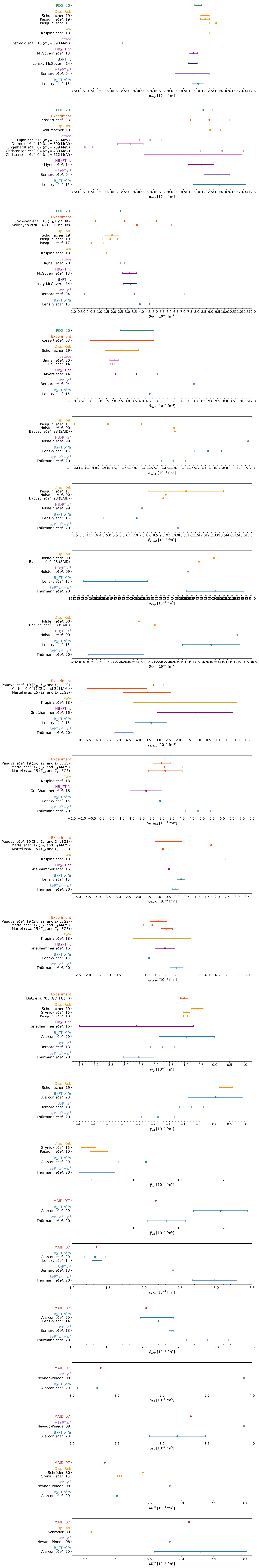}
\caption{Summary for the longitudinal-transverse polarizability of the proton $\delta_{LTp}$ (upper panel) and neutron $\delta_{LTn}$ (lower panel). Theoretical predictions from chiral EFT are compared to the MAID unitary isobar model. \label{LTPolarizabilities}}
\end{figure}

There are two prominent counting schemes for $\chi$PT with explicit inclusion of the $\Delta(1232)$.
For simplicity, they both deduce a single expansion parameter from the two involved small mass scales: $\epsilon=m_\pi/\Lambda_{\chi\mathrm{SB}}$ and 
$\delta=\varDelta/\Lambda_{\chi\mathrm{SB}}$. In the $\eps$-expansion (small-scale expansion) it is assumed that $\eps \sim \delta$ \cite{Hemmert:1996xg}, while in the $\delta$-expansion one assumes $\eps \sim \delta^2$ with $\eps \ll \delta$ \cite{Pascalutsa:2003aa}. In this way, the $\delta$-expansion defines a hierarchy between the two mass scales. Consequently, it defines two regimes where the $\Delta(1232)$ contributions need to be counted differently:
\begin{itemize}[leftmargin=*,labelsep=5.8mm]
\item low-energy region: $p \sim m_\pi$;
\item resonance region: $p\sim \varDelta$.
\end{itemize}
This makes sense since the $\Delta(1232)$ is expected to be suppressed at low energies and dominating in the resonance region. The chiral order $\mathcal{O}(p^{n_\delta})$ of a Feynman diagram with $N_\text{1$\Delta$R}$ ($N_\text{1$\Delta$I}$)  one-$\Delta$-reducible (one-$\Delta$-irreducible) propagators is in the $\delta$-expansion defined as:
\beq
n_\delta=\begin{cases}n-\nicefrac{1}{2}\,N_\Delta\,,&p\sim m_\pi,\\
n-3N_\text{1$\Delta$R}-N_\text{1$\Delta$I}\,,&p\sim\Delta,\eqlab{deltacounting}
\end{cases}
\eeq
where
\beq
N_\Delta=N_\text{1$\Delta$R}+N_\text{1$\Delta$I}.
\eeq
An extensive review on the electromagnetic excitation of the $\Delta$(1232) resonance with more details on the formulation of the extended $\chi$PT framework and the chiral expansion in the resonance region can be found in Ref.~\cite{Pascalutsa:2006up}. 
As we will see in Section \ref{results}, B$\chi$PT calculations based on the
$\eps$ \cite{Bernard:2012hb} and the
$\de$ \cite{Lensky:2014dda,Alarcon:2020icz} counting schemes give significantly different predictions for the longitudinal-transverse polarizability of the proton shown in Figures~\ref{LTPolarizabilities} (upper panel) and \ref{Fig:deltaLTQ2}.

\subsection{Low-Energy Constants and Predictive Orders} \label{LECsec}

At any given order in the chiral expansion, the divergencies of the EFT
are absorbed by renormalization of a finite number of LECs.
To match $\chi$PT to QCD as the fundamental theory of the strong interaction, the renormalized LECs need to be fitted to experimental or lattice data. Thereby, it is important that the LECs are constrained to be of {\it natural size}. Take for instance the
fifth-order forward spin polarizability (in units of $10^{-4}$~fm$^6$) \cite{Alarcon:2020icz}:
\begin{subequations}
\begin{align}
&\bar\gamma_{0p} =1.12(30)\approx 2.08\,\mbox{($\pi$N loop)} -0.96\,\mbox{($\Delta$ exchange)} -0.01\, \mbox{($\pi\Delta$ loop)}, \label{Eq:gamma0ProtonRealPoint}\\
&\bar\gamma_{0n} =1.95 (30) \approx 2.92\,\mbox{($\pi$N loop)}-0.96\,\mbox{($\Delta$ exchange)} -0.01\, \mbox{($\pi\Delta$ loop)},
\end{align}
\end{subequations}
also shown in Figure \ref{BarGamma0Pol}.
The next-to-leading-order effect of the $\Delta(1232)$ is two to three times smaller than the leading-order effect of the pion cloud. This is consistent with estimates from power counting, according to which each subleading order is expected to be suppressed with respect to the previous one by a factor of $\sim\varDelta/M_N\sim 1/3$.
Therefore, implementing this {\it naturalness} allows to estimate the uncertainty due to neglect of higher-order effects.

The LECs entering a next-to-next-to-leading-order B$\chi$PT calculation of low-energy CS in the $\delta$-expansion are $f_\pi$, $g_A$, $h_A$, $g_M$, $g_E$ and $g_C$.\footnote{Note that the $g_E$ and $g_C$ couplings of the $N$-to-$\Delta$ transition would be strictly speaking of higher order.} They are listed in Table \ref{LECtable} together with the experiments used to constrain their values. As one can see, B$\chi$PT has 
``predictive power'' for low-energy CS up to and including $\mathcal{O}(p^4/\varDelta)$ because all relevant LECs are matched to processes other than CS.\footnote{Note that $\mathcal{O}(p^4/\varDelta)$ corresponds to $\mathcal{O}(p^{7/2})$, cf.\ \Eqref{deltacounting} with $p
^{1/2} \sim \varDelta $ or $p \sim m_\pi$.} This makes $\chi$PT the perfect tool to study the low-energy structure of the nucleon as encoded in CS and the associated polarizabilities. Starting from $\mathcal{O}(p^4)$, LECs need to be fitted to the CS process as well, for instance through the Baldin sum rule, as done in  Refs.~\cite{McGovern:2012ew,Griesshammer:2012we,Lensky:2014efa,Myers:2014ace,Myers:2015aba,Griesshammer:2015ahu,Alarcon:2020wjg}.

\subsection{Heavy-Baryon Expansion}\label{secHB}

The theory of HB$\chi$PT  was first introduced in Ref.~\cite{Jenkins:1990jv}, and later applied to CS and polarizabilities \cite{Butler:1992ci}, including also the effect of the $\Delta(1232)$ \cite{Bernard:1995dp,Hemmert:1996rw,Hildebrandt:2003fm,Griesshammer:2012we,Kao:2002cp,Kao:2003jd,Nevado:2007dd}. The results of HB$\chi$PT can be recovered from the B$\chi$PT results by expanding in powers of the inverse nucleon mass. HB$\chi$PT calculations tend to fail in describing the $Q
^2$ evolution of the generalized nucleon polarizabilities \cite{Alarcon:2020icz,Alarcon:2020wjg}. 
Also for the polarizabilities at the real-photon point ($Q^2=0$) the heavy-baryon expansion can give significantly different predictions. Consider for instance the nucleon dipole polarizabilities. The B$\chi$PT prediction (in units of $10^{-4}\,$fm$^3$) \cite{Lensky:2015awa}:
\begin{subequations}
\begin{eqnarray}
\eqlab{alChPT}
 \alpha_{E1p}&=& 6.9\,\mbox{($\pi$N loop)} -0.1\,\mbox{($\Delta$ exchange)} + 4.4\, \mbox{($\pi\Delta$ loop)} = 11.2\pm 0.7,\\
\beta_{M1p}&=&-1.8\,\mbox{($\pi$N loop)} + 7.1\,\mbox{($\Delta$ exchange)}-1.4\, \mbox{($\pi\Delta$ loop)} =3.9\pm 0.7,
\end{eqnarray}
\end{subequations}
is in good agreement with empirical evaluations, see Figures~\ref{alphaE1Pol} and \ref{betaM1Pol}.
In HB$\chi$PT, however, the $\De(1232)$ contributions
to the nucleon polarizabilities turn out to be large \cite{Hemmert:1996rw}
and need to be canceled by promoting the higher-order [$\mathcal{O}(p^4)$] counterterms $\delta \alpha$ and $\delta \beta$ (in units of $10^{-4}\,$fm$^3$) \cite{Hildebrandt:2003fm}:
\begin{subequations}
\eqlab{abHBChPT}
\bea 
  \alpha_{E1p}(\mbox{HB}) &=& 11.87 \,\mbox{($\pi$N loop)} + 0 \,\mbox{($\Delta$ exch.)} + 5.09 \, \mbox{($\pi\Delta$ loop)} -(5.92\pm 1.36)\, \mbox{($\delta \alpha$)} \nn\\
  &=& 11.04\pm1.36\,,\\
 \beta_{M1p}(\mbox{HB})&=& 1.25 \,\mbox{($\pi$N loop)}  
 + (11.33\pm 0.70) \,\mbox{($\Delta$ exch.)} +0.86\,  \mbox{($\pi\Delta$ loop)}\nn\\
 &&-(10.68\pm 1.17)\, \mbox{($\delta \beta$)}\qquad\nn \\
  &=&  2.76\mp 1.36,
\eea 
\end{subequations}
at the expense of 
violating the naturalness requirement, see also Ref.~\cite{Griesshammer:2012we}. This can be seen from the dimensionless LECs associated to $\delta \alpha$ and $\delta \beta$, $g_{117}=18.82\pm0.79$ and $g_{118}=-6.05\mp 0.66$ \cite{Hildebrandt:2003fm}, that should be of $\mathcal{O}(1)$ to be consistent with estimates from power counting. This problem is
discussed at length in Refs.~\cite{Lensky:2009uv,Hall:2012iw}.

\begin{figure}[t]
\begin{center}
\includegraphics[width=0.9\columnwidth]{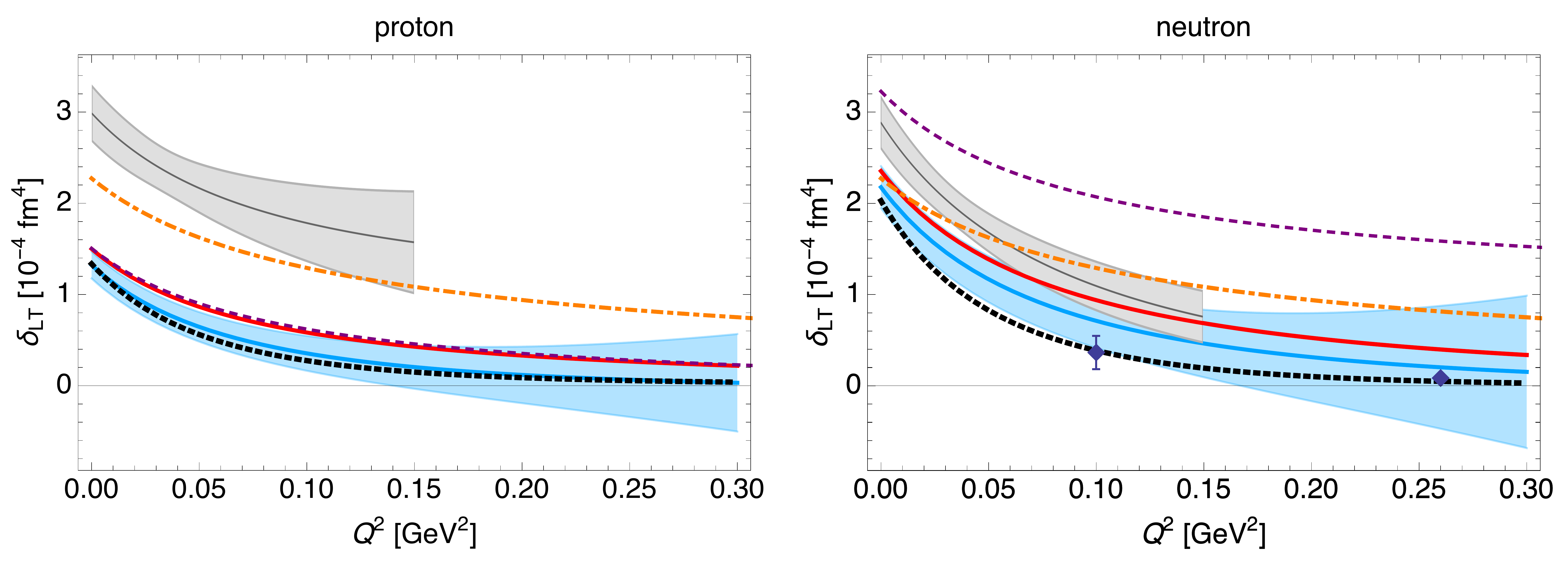}
\caption{Longitudinal-transverse spin polarizability, \Eqref{dLTgen}, for the proton (left) and neutron (right) as function of $Q^2$. The black dotted line is the MAID model~\cite{Drechsel:2000ct,Drechsel:1998hk}; note that for the proton we use the updated estimate from Ref.~\cite{Drechsel:2002ar} obtained using the $\pi,\eta,\pi\pi$ channels. The red line shows the leading-order B$\chi$PT result. The blue band is the $\mathcal{O}(p
^4/\Delta)$ B$\chi$PT result from  Ref.~\cite{Alarcon:2020icz}. The gray band is the $\mathcal{O}(\epsilon^3+p
^4)$ B$\chi$PT result from Ref.~\cite{Thurmann:2020mog}.
 The orange dot-dashed and purple short-dashed lines are the $\mathcal{O}(p^3)$ and $\mathcal{O}(p^4)$ HB results from Ref.~\cite{Kao:2002cp}. The experimental points for the neutron are from Ref.~\cite{Amarian:2004yf} (blue diamonds).
 \label{Fig:deltaLTQ2}}
\end{center}
\end{figure}

\section{Compton Scattering Formalism} \label{CSintro}

The CS process, shown in Figure~\ref{CSfigure}, gives the most direct access to the nucleon polarizabilities. Of interest are the following kinematic regimes, described by the four-momenta of incoming (outgoing) photons $q(q')$ and nucleons $p(p')$:
\begin{itemize}[leftmargin=*,labelsep=5.8mm]
\item	Real Compton scattering (RCS): $q^2=q^{\prime\,2}=0$;
\item	Virtual Compton scattering (VCS): $q^2=-Q^2<0$ and $q^{\prime\,2}=0$;
\item	Forward doubly-virtual Compton scattering (VVCS): $q=q'$ (thus $p=p'$) and $q^2=-Q^2<0$.
\end{itemize}
In general kinematics ($p^2=p^{\prime\,2}=M_N^2$, $q^2 \neq q^{\prime\,2}$), the CS amplitude can be described by $18$ independent tensor structures. For VCS one needs $12$ independent tensor structures; for RCS one needs $6$ independent tensor structures \cite{Hearn:1962zz,Babusci:1998ww}. In the forward limit, this reduces to $4$ independent tensor structures for virtual photons and $2$ independent tensor structures for real photons.

\begin{figure}[t]
\centering
\hspace{-2cm}\includegraphics[width=0.9\columnwidth]{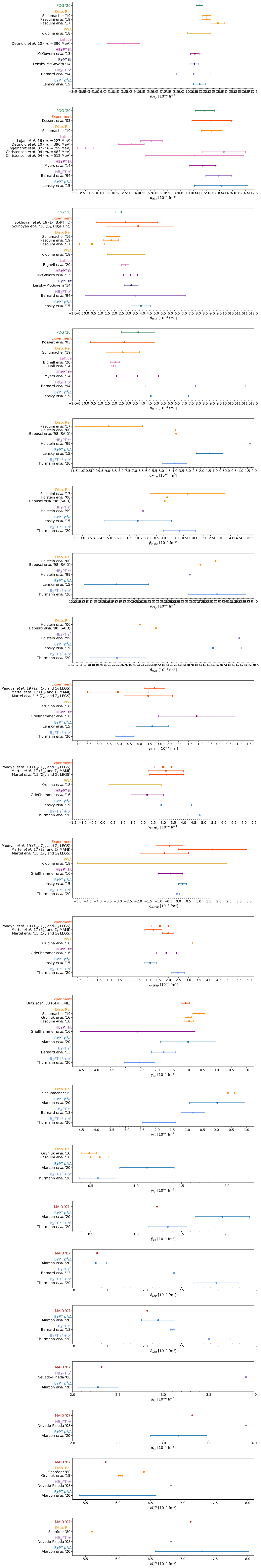}
\caption{Summary for the fifth-order forward spin polarizability of the proton $\bar \ga_{0p}$ (upper panel) and neutron $\bar \ga_{0n}$ (lower panel). Theoretical predictions from chiral EFT are compared to empirical evaluations of the fifth-order forward spin polarizability sum rule \eref{barg0gen} at the real-photon point and the MAID unitary isobar model.\label{BarGamma0Pol}}
\end{figure}

Splitting into spin-independent (symmetric) and spin-dependent (antisymmetric) parts, the forward VVCS decomposes into four scalar amplitudes $T_i(\nu,Q^2)$ and $S_i(\nu,Q^2)$:
\begin{subequations}
\beq
T^{\mu \nu} (q,p) = \left[T^{\mu \nu}_S+T^{\mu \nu}_A\right] (q,p) ,
\eeq
with
\bea
T^{\mu \nu}_S(q,p) & = & -g^{\mu\nu}\,
T_1(\nu, Q^2)  +\frac{p^{\mu} p^{\nu} }{M_N^2} \, T_2(\nu, Q^2), \eqlab{VVCS_TS}\\
T^{\mu \nu}_A (q,p) & = &-\frac{1}{M_N}\gamma^{\mu \nu \al} q_\al \,S_1(\nu, Q^2) +
\frac{Q^2}{M_N^2}  \gamma^{\mu\nu} S_2(\nu, Q^2),
\eea
\end{subequations}
with $\nu$ the photon lab-frame energy, $Q^2$ the photon virtuality, and terms which vanish upon contraction with the photon polarization vectors omitted. For real photons, the following two scalar amplitudes  survive: 
\beq
f(\nu)=\frac{1}{4\pi}T_1(\nu,0), \qquad g(\nu)=\frac{\nu }{4\pi M_N} S_1(\nu,0).
\eeq
Constraints relating the different kinematic regimes (RCS, VCS and forward VVCS) are discussed in Refs.~\cite{Lensky:2017bwi} and \cite{Pascalutsa:2014zna,Lensky:2017dlc} for the unpolarized and polarized CS, respectively. Here, the focus is on RCS and forward VVCS.

\begin{figure}[t]
\centering
\hspace{-1.5cm}
\includegraphics[width=0.9\columnwidth]{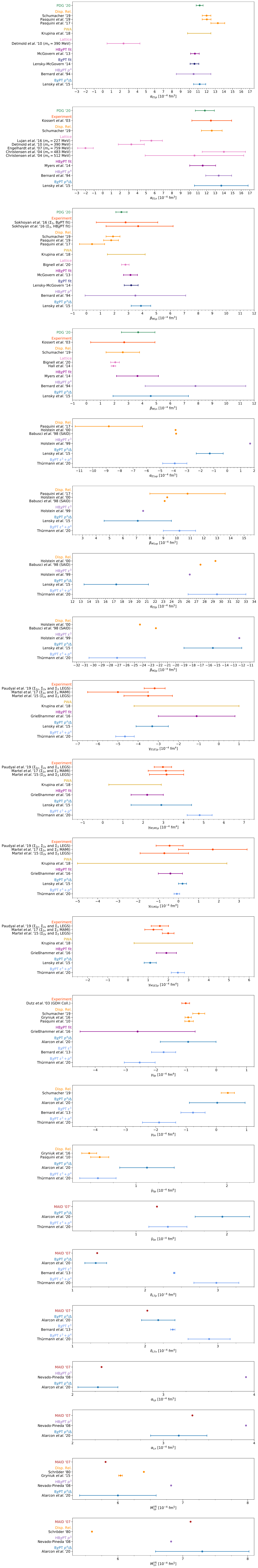}
\caption{Summary for the quadrupole polarizabilities $\al_{E2p}$ and $\be_{M2p}$ of the proton. Theoretical predictions from chiral EFT are compared with extractions based on CS data.\label{QuadrupolePol}}
\end{figure} 

The off-forward RCS is conveniently described by the covariant decomposition~\cite{Pascalutsa:2003aa}: 
\begin{subequations}
\eqlab{CovarAmp}
 \beq
 \bar u^{\,\prime} (\veps' \cdot T \cdot \veps)  u  = 4\pi \al\,\hat \scA^T (s,t)\, \bar u^{\,\prime} \hat \scO^{\mu \nu}  u \,  \mathcal{E}_{\mu}^{\prime }  \mathcal{E}_{\nu},
 \eeq
 with the overcomplete set of $8$ tensors:
  \bea
&& \hat \scA(s,t) = \big\{\scA_1, \, \cdots, \, \scA_8 \big\} (s,t), \\
&& \hat \scO^{\mu\nu} = \big\{ -g^{\mu \nu}, \; q^{\mu} q^{\prime\,\nu},\; 
-\gamma^{\mu \nu},\; g^{\mu \nu} (q' \cdot \gamma \cdot q),\; q^{\mu} q'_{\alpha} \gamma^{\alpha \nu}-\gamma^{\alpha \mu} q_{\alpha} q'^{\nu},\;q^{\mu} q_{\alpha} \gamma^{\alpha \nu}-\gamma^{\alpha \mu} q'_{\alpha} q'^{\nu},\nn \\  
&& \qquad\qquad 
q^{\mu} q^{\prime\,\nu}(q' \cdot \gamma \cdot q) ,\; 
- i \gamma_5 \epsilon^{\mu \nu \alpha \beta} q'_{\alpha} q_{\beta}\big\}, \\
&& \mathcal{E}_{\mu}  = \veps_{\mu} - \frac{P \cdot \veps}{P \cdot q} \, q_{\mu},
\quad \mathcal{E}_{\mu}'  = \veps_{\mu}' - \frac{P \cdot \veps'}{P \cdot q} \, q_{\mu}',
\quad P_\mu = \half (p + p')_\mu, \quad P\cdot q=P\cdot q',
 \eqlab{epsilon_b}
\eea 
\end{subequations}
and the incoming (outgoing) photon polarization vector $\veps^{(\prime)}$ and Dirac spinor $u^{(\prime)}$. Alternatively, one can choose the non-covariant decomposition with the minimal set of $6$ tensors:
\begin{subequations}
\eqlab{helAmp}
\beq
\bar u^{\,\prime} (\veps' \cdot T \cdot \veps)  u =  8\pi \al M_N  \,  \hat A^T (s, t)\, 
 \chi^{\,\prime} \veps_i' \, \hat O_{ij}\,  \veps_j\,  \chi, 
\eeq 
with the incoming (outgoing) Pauli spinor $\chi^{(\prime)}$ and the scalar complex amplitudes \cite{Hemmert:1997tj}:
\bea
&& \hat A(s,t) = \big\{A_1, \, \cdots, \, A_6 \big\} (s,t), \\
&& \hat O_{ij} = \big\{ \de_{ij}, \, n_i n_j',\, i \eps_{ijk} \si_k,\, \de_{ij}
 i \eps_{klm} \si_k n_l' n_m , \,i \eps_{k lm}\si_k ( \de_{il} n_m n_j' -  
 \de_{jl} n_i n_m'),\nn\\
 &&\qquad\qquad i \eps_{k lm}\si_k ( \de_{il} n_m' n_j' -  
 \de_{jl} n_i n_m) \big\},\qquad\qquad
\eea
\end{subequations}
where $\vec{n}^{(\prime)}$ is the direction of the incoming (outgoing) photon, $\sigma_k$ are the Pauli matrices and $\delta_{ij}$ is the Kronecker delta. The scalar amplitudes $\scA_{1,\dots, 8}$ are related to the scalar amplitudes $A_{1,\dots, 6}$ in the following way
\cite{McGovern:2012ew}: 
\begin{subequations}
\eqlab{eq:relations_Ai_covarAi}
\bea
A_1& =& \frac{\eps_B}{M_N} \scA_1 + \frac{\omega_B t }{2 M_N} \scA_4, \\
A_2 &=& \frac{\eps_B \w_B^2 }{M_N} \scA_2 + \frac{\omega_B^3}{M_N} \left(
\scA_5 + \scA_6  -  \half t \scA_7 \right),  \\
A_3 &=&  \frac{\eps_B }{M_N} \scA_3 - \frac{M_N^2 \eta\, t}{4M_N^2-t} 
\left(\frac{\scA_5 + \scA_6 }{2M_N(\eps_B +M_N)} - \scA_7\right) - 
\frac{\omega_B t }{2 M_N} \scA_8, \\
A_4 &=& \w_B^2 \scA_4, \\
A_5 &=& \w_B^2 \scA_5 +  \frac{\w_B^2}{2M_N (\eps_B + M_N)} \Big[\half \scA_3 
+ \frac{M_N^2\eta}{4M_N^2-t}  \left(\scA_5 + \scA_6 \right) \Big] 
-\w_B^2 (\omega_B^2 + \half t) \scA_7 + \frac{\omega_B^3}{2 M_N} \scA_8,  \qquad\\
A_6 &=& \w_B^2 \scA_6 -  \frac{\w_B^2}{2M_N (\eps_B + M_N)} \Big[\half \scA_3 
+ \frac{M_N^2\eta}{4M_N^2-t}  \left(\scA_5 + \scA_6 \right) \Big] + \omega_B^4 \scA_7 - \frac{\omega_B^3}{2 M_N} \scA_8, 
\eea
\end{subequations}
where 
\begin{subequations}
\bea
\w_B &=& \frac{s-u}{2\sqrt{4M_N^2 -  t}},\\
\eps_B &=& \frac12 \sqrt{4M_N^2 -  t}.
\eea
\end{subequations}
are the nucleon and photon energies in the 
Breit frame ($\vec{p}'=-\vec{p}\,$),  
\beq
\eta=\frac{M_N^4-su}{M_N^2},
\eeq
and $s$, $t$, $u$ are the usual Mandelstam variables.

\begin{figure}[t]
\centering
\hspace{-1.7cm}\includegraphics[width=0.9\columnwidth]{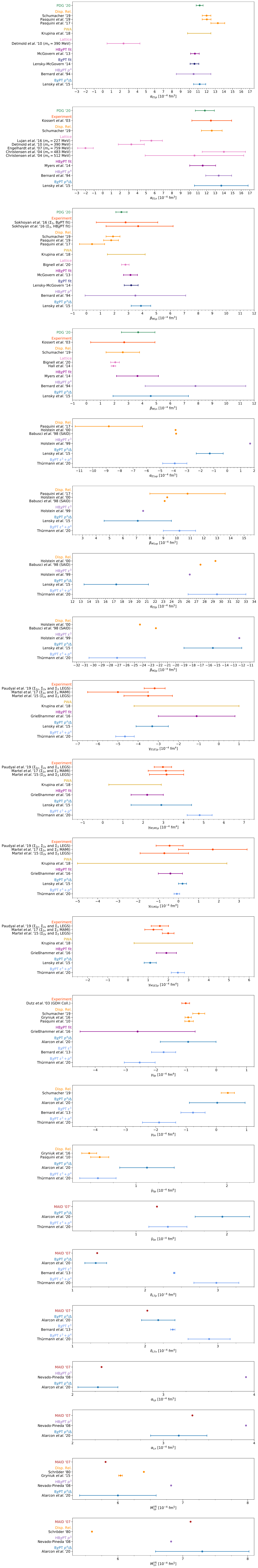}
\caption{Summary for the dispersive polarizabilities  of the proton, $\al_{E1\nu p}$ and $\be_{M1\nu p}$. Theoretical predictions from chiral EFT are compared with extractions based on CS data. Note that Pasquini et al.\ '17 \cite{Pasquini:2017ehj} presented the first extraction of the dispersive polarizabilities from proton RCS data below pion-production threshold.\label{DispersivePol}}
\end{figure}

\begin{figure}[h!]
\centering
\includegraphics[width=0.9\columnwidth]{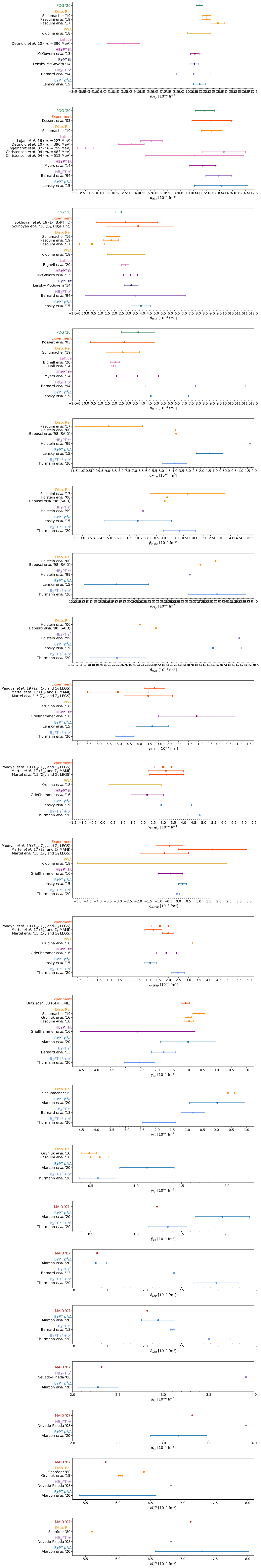}
\caption{Summary for the lowest-order spin polarizabilities $\ga_{E1E1p}$, $\ga_{M1M1p}$, $\ga_{E1M2p}$ and $\ga_{M1E2p}$ of the proton. Theoretical predictions from chiral EFT are compared with extractions based on CS data. The experimental results are combinations of different beam asymmetry and double-polarization observable measurements at MAMI and LEGS: $\Sigma_{2x}$ \cite{Martel:2017pln,Martel:2014pba}, $\Sigma_{2z}$ \cite{Paudyal:2019mee} and $\Si_3$ \cite{Sokhoyan:2016yrc,Blanpied:2001ae}.  Krupina et al.~\cite{Krupina:2017pgr} performed a  PWA of proton RCS data below pion-production threshold.\label{ForwardSpinPolarizabilities}}
\end{figure}

According to the low-energy theorem of Low \cite{ Low:1954kd}, Gell-Mann and Goldberger \cite{GellMann:1954kc}, the leading terms in a low-energy expansion of the RCS amplitudes are determined by charge, mass and anomalous magnetic moment of the nucleon. At higher orders in the low-energy expansion various polarizabilities emerge. 
The low-energy expansion of the non-Born RCS amplitudes (denoted by an overline, e.g., $\bar A_{1,\dots,6}$) reads as:
\begin{subequations}
\label{eq:anb}
\bea
\al \bar A_1(\w_B ,t) &=&  \omega_B^2 \big[ \alpha_{E1}+ \beta_{M1} + 
\omega_B^2\, (\alpha_{E1\nu}+ \beta_{M1\nu})\, \big] + \half t 
\big( \beta_{M1} + \w_B^2 \beta_{M1\nu} \, \big) \nn\\
&&+\;  \w_B^4 \boxfrac{1}{12} ( \alpha_{E2}+\beta_{M2}) 
+\half t (4\w_B^2+t)\boxfrac{1}{12} \beta_{M2}+ \mathcal{O}(\omega_B ^6), \\ 
\al \bar A_2 (\w_B ,t)&=&   - \w_B^2 \big( \beta_{M1} + \w_B^2 \beta_{M1\nu}
\big) +
\w_B^4 \boxfrac{1}{12} ( \alpha_{E2}-\beta_{M2}) -
t \w_B^2\boxfrac{1}{12} \beta_{M2}+ \mathcal{O}(\omega_B^6),\\
\al \bar A_3(\w_B ,t) &=& -  \omega_B^3 \, \big[\gamma_{E1E1}+\gamma_{E1M2} + z\, (\gamma_{M1E2}+\gamma_{M1M1})  \big] + \mathcal{O}(\omega_B^5),  \\ 
\al \bar A_4(\w_B ,t) &=&  \omega_B^3\, (\gamma_{M1E2}-\gamma_{M1M1})   + \mathcal{O}(\omega_B^5),   \\
\al \bar A_5(\w_B ,t) &=&   \omega_B^3 \, \gamma_{M1M1}+ \mathcal{O}(\omega_B^5),\\
\al \bar A_6 (\w_B ,t) &=& \omega_B^3 \,  \gamma_{E1M2}+ \mathcal{O}(\omega_B^5),
\eea
\end{subequations}
with $z=\cos \theta_B = 1+t/2\omega_B^2$ and $\theta_B$ the scattering angle in the Breit frame. The coefficients are given in terms of static nucleon polarizabilities: electric dipole ($\al_{E1}$),  magnetic dipole ($\beta_{M1}$), quadrupole ($\al_{E2}$, $\beta_{M2}$), dispersive ($\al_{E1\nu}$, $\beta_{M1\nu}$), and lowest-order spin polarizabilities ($\ga_{E1E1}$, $\ga_{M1M1}$, $\ga_{E1M2}$ and $\ga_{M1E2}$), see Figures~\ref{alphaE1Pol}, \ref{betaM1Pol}, \ref{QuadrupolePol}, \ref{DispersivePol} and \ref{ForwardSpinPolarizabilities}, respectively. The latter combine into the forward (see Figure~\ref{Gamma0Pol}) and backward spin polarizabilities:
\begin{subequations}
\bea
\ga_0 &=& - \ga_{E1E1}- \ga_{M1M1}-\ga_{E1M2}
-\ga_{M1E2},\\
\ga_\pi &=& - \ga_{E1E1}+ \ga_{M1M1}-\ga_{E1M2}
+\ga_{M1E2}.
\eea
\end{subequations}

Studying the forward RCS and VVCS is of advantage because of their accessibility through sum rules. 
Based on the general principles of causality, unitarity and crossing symmetry, the forward VVCS amplitudes can be expressed in terms of the nucleon structure functions by means of dispersion relations and the optical theorem \cite{Drechsel:2002ar}:
\begin{subequations}
\eqlab{genDRs}
\bea 
T_1 ( \nu, Q^2) &=&T_1(0,Q^2) + \frac{32\pi\al M_N\nu^2}{Q^4} \int_{0}^1 
\!\dd x \,
\frac{x\, f_1 (x, Q^2)}{1 - x^2 (\nu/\nu_{\mathrm{el}})^2 - i 0^+} \eqlab{T1dr}\\ 
&=& T_1(0,Q^2) + \frac{2\nu^2}{\pi } \int_{\nu_{\mathrm{el}}}^\infty\! \frac{\dd \nu'}{\nu'} \, 
\frac{\sqrt{\nu^{\prime\,2}+Q^2}\,\si_T ( \nu', Q^2)}{\nu^{\prime\,2} -\nu^2 - i 0^+}\,,\nn\\
T_2 ( \nu, Q^2) &=& \frac{16\pi\al M_N}{Q^2} \int_{0}^1 
\!\dd x\, 
\frac{f_2 (x, Q^2)}{1 - x^2 (\nu/\nu_{\mathrm{el}})^2  - i 0^+} \eqlab{T2dr}\\
&=&\frac{2Q^2}{\pi} \int_{\nu_{\mathrm{el}}}^\infty\! \dd \nu' \, 
\frac{\nu^{\prime}\,[ \si_T+\si_L] ( \nu', Q^2)}{\sqrt{\nu^{\prime\,2}+Q^2}(\nu^{\prime\,2} -\nu^2- i 0^+)} ,\nn\\
S_1 ( \nu, Q^2) &=&  \frac{16\pi\al M_N}{Q^2} \int_{0}^1 
\!\dd x\, 
\frac{g_1 (x, Q^2)}{1 - x^2 (\nu/\nu_{\mathrm{el}})^2  - i 0^+} \eqlab{S1DR}\\
&=&\frac{2M_N}{\pi} \int_{\nu_{\mathrm{el}}}^\infty \!\dd \nu' \, 
\frac{\nu^{\prime\,2}\big[ \frac{Q}{\nu'}\si_{LT}+\si_{TT}\big] ( \nu', Q^2)}{\sqrt{\nu^{\prime\,2}+Q^2}(\nu^{\prime\,2} -\nu^2- i 0^+)}\nn,\\
\nu S_2 ( \nu, Q^2) &=& \frac{16\pi\al M_N^2}{Q^2} \int_{0}^1 \!\dd x\, 
\frac{g_2 (x, Q^2)}{1 - x^2 (\nu/\nu_{\mathrm{el}})^2  - i 0^+}\eqlab{nuS2}  \\
&=& \frac{2M_N^2}{\pi} \int_{\nu_{\mathrm{el}}}^\infty \!\dd \nu' \, 
\frac{ \nu^{\prime\,2}\big[ \frac{\nu'}{Q}\si_{LT}-\si_{TT}\big] ( \nu', Q^2)}{\sqrt{\nu^{\prime\,2}+Q^2}(\nu^{\prime\,2} -\nu^2- i 0^+)},\nn
\eea 
\end{subequations}
with $\nu_{\mathrm{el}}=Q^2/2M_N$ the elastic threshold. Note that the structure functions $f_1$, $f_2$, $g_1$ and $g_2$ are functions of the Bjorken variable $x=\nu_{\mathrm{el}}/\nu$ and the photon virtuality $Q^2$. They are related to the photoabsorption cross sections $\sigma_T$, $\sigma_L$, $\sigma_{TT}$ and $\sigma_{LT}$ measured in electroproduction, defined here with the photon flux factor $K(\nu,Q
^2)=\sqrt{\nu^2+Q^2}$ \cite{Gilman:1967sn}.

Performing low-energy expansions of the relativistic CS amplitudes \cite{Drechsel:1998zm,Drechsel:2002ar,Pascalutsa:2014zna}
and combining these with dispersion relations and the optical theorem leads to various sum rules for the polarizabilities. A famous sum-rule example is the Baldin sum rule \cite{Baldin:1960}, allowing for a precise data-driven evaluation of the sum of electric and magnetic dipole polarizabilities, cf.\ Eqs.\ (\ref{eq:BaldinSumRule}) and (\ref{BaldinSRresults}).
It follows from the $\nu^2$ term in the low-energy expansion of the RCS amplitude $f(\nu)$:
\beq
f(\nu)=-
\, \frac{ Z^2 \al}{M_N}+ \left[\alpha_{E1}+\beta_{M1}\right]\,\nu^2+\left[\alpha_{E1\nu} + \beta_{M1\nu} + \nicefrac{1}{12}\,(\alpha_{E2} + \beta_{M2}) \right]\,\nu^4
+\mathcal{O}(\nu^6)\eqlab{T1LEX}.
\eeq
The extension of the Baldin sum rule to finite momentum-transfers \cite{Drechsel:2002ar},
\beq
\left[\al_{E1}+\beta_{M1}\right](Q^2)=\frac{1}{2 \pi^2} \int_{\nu_0}^\infty \! \dd\nu\,\sqrt{1+\frac{Q^2}{\nu^2}}\, \frac{\sigma_T (\nu,Q^2)}{\nu^2},\eqlab{alphabetaf1}
\eeq
defines the $Q^2$ dependent sum of generalized dipole polarizabilities. Be aware that while the definitions of the polarizabilities in the real-photon limit are unambiguous, the generalized polarizabilities defined in VCS and forward VVCS can differ. As an example, one can consider the magnetic dipole polarizability $\beta_{M1}(Q^2)$, which for VCS is defined in Eq.~(B2b) of Ref.~\cite{Lensky:2017bwi}, and for forward VVCS could be defined either by generalizing the non-Born part of the subtraction function 
\beq
\eqlab{subtractionfunction}
\frac{ \ol T_1(0,Q
^2)}{4\pi}= \beta_{M1}Q^2+\mathcal{O}(Q^4), 
\eeq
but is usually understood as part of the generalized Baldin sum rule \eref{alphabetaf1}. 
A recent measurement of the generalized $\al_{E1}(Q^2)$ and $\beta_{M1}(Q^2)$ polarizabilities from VCS by the A1 Collaboration can be found in Ref.~\cite{Bericic:2019faq}.

The generalized fourth-order Baldin sum rule is defined as:
\beq
\eqlab{SumRuleBaldin4}
 M_1^{(4)}(Q^2)=
\frac{1}{2 \pi^2} \, \int_{\nu_0}^{\infty}\, \mathrm{d}\nu \,\sqrt{1+\frac{Q^2}{\nu^{2}}}\, \frac{\sigma_T(\nu,Q^2)}{\nu^{4} }.
\eeq
It differs from the generalized Baldin sum rule \eref{alphabetaf1} by the energy weighting of the total photoabsorption cross section $\sigma_T$ in the sum rule integral.
In the real-photon limit, it is related to a linear combination of the dispersive and quadrupole polarizabilities given by the $\nu^4$ term in \Eqref{T1LEX} \cite{Guiasu:1978dz,Holstein:1999uu}:
\beq
M_1^{(4)}(0)=\alpha_{E1 \nu} + \beta_{M1 \nu} + \frac{1}{12} (\alpha_{E2} + \beta_{M2}),
\eeq
see  Figure~\ref{Baldin4Pol}.

Similarly, the low-energy expansion of the RCS amplitude $g(\nu)$:
\beq
g(\nu)=-\frac{\al \kappa_N^2}{2M_N^2}\,\nu+\ga_0\, \nu^3+\bar \gamma_0 \,\nu^5+\mathcal{O}(\nu^6),
\eeq
allows to express the anomalous magnetic moment of the nucleon ($\kappa_p\sim1.79, \kappa_n\sim-1.91$) and the forward spin polarizabilities as sum rule integrals over the helicity-difference photoabsorption cross section $\sigma_{TT}$, cf.\ \Eqref{S1DR}. The Gerasimov--Drell--Hearn sum rule \cite{Gerasimov:1965et,Drell:1966jv}:
\beq
-\frac{\alpha}{2 M_N^2 \kappa_N^2} = \frac{1}{2 \pi^2} \int_{\nu_0}^\infty \! \dd\nu\,\frac{\sigma_{TT} (\nu)}{\nu}, \label{Eq:GDH-SumRule}
\eeq
has been experimentally verified for the nucleon by MAMI (Mainz) and ELSA (Bonn)~\cite{Ahrens:2001qt, Helbing:2002eg}.  The same cross section input can be used to evaluate the forward spin polarizabilities at the real-photon point, cf.\ Figures~\ref{Gamma0Pol} and \ref{BarGamma0Pol}. Considering the extension to finite momentum-transfers, the generalized forward spin polarizability reads \cite{Drechsel:2002ar}: 
\beq
\eqlab{g0gen}
\gamma_0(Q^2)=\frac{1}{2 \pi^2} \int_{\nu_0}^\infty \! \dd\nu\,\sqrt{1+\frac{Q^2}{\nu^2}} \,\frac{\sigma_{TT} (\nu,Q^2)}{\nu^3},
\eeq
while the fifth-order generalized forward spin polarizability sum rule is given by: \beq
\eqlab{barg0gen}
\bar\gamma_0 (Q^2)= \frac{1}{2 \pi^2} \int_{\nu_0}^\infty \! \dd\nu\,\sqrt{1+\frac{Q^2}{\nu^2}} \,\frac{\sigma_{TT} (\nu,Q^2)}{\nu^5},
\eeq
see Figure \ref{Fig:gamma0plot} upper and lower panel, respectively.

The polarizabilities involving longitudinal photon polarizations are absent from RCS. They are given as sum rule integrals over the longitudinal photoabsorption cross section $\sigma_L$, e.g., the longitudinal polarizability \cite{Lensky:2014dda}:
\beq
\eqlab{SumruleaL}
\al_{L}(Q^2)=\frac{1}{2 \pi^2} \int_{\nu_0}^\infty\! \dd\nu\,\sqrt{1+\frac{Q^2}{\nu^{2}}}\,            \frac{\sigma_L (\nu,Q^2)}{Q^2\, \nu^{2}},
\eeq
cf.\ Figure \ref{LongitudinalPolarizabilities}, 
and the longitudinal-transverse cross section $\sigma_{LT}$, e.g., the longitudinal-transverse polarizability \cite{Drechsel:2002ar}:
\beq
\eqlab{dLTgen}
\delta_{LT}(Q^2)=\frac{1}{2 \pi^2} \int_{\nu_0}^\infty \! \dd\nu\,\sqrt{1+\frac{Q^2}{\nu^{2}}}\, \frac{\sigma_{LT} (\nu,Q^2)}{Q\,\nu^2},
\eeq
see Figures~ \ref{LTPolarizabilities}, and \ref{Fig:deltaLTQ2}.

\section{Nucleon Polarizabilities}\label{results}

In the following, I want to discuss the nucleon polarizabilities, focusing on new empirical results from the last five years and comparisons to $\chi$PT predictions.
References quoted in the summary figures are: PDG \cite{Zyla:2020}, MAID \cite{MAID}, experiments \cite{Paudyal:2019mee,Martel:2017pln,Sokhoyan:2016yrc,Martel:2014pba,Dutz:2003mm,Kossert:2002ws}, dispersion relations \cite{Schumacher:2019ikn, Pasquini:2019nnx,Pasquini:2017ehj,Gryniuk:2016gnm,Pasquini:2010zr,Holstein:1999uu,Babusci:1998ww,Schroder:1977sn}, PWA \cite{Krupina:2017pgr}, lattice QCD \cite{Bignell:2020xkf,Lujan:2016ffj,Hall:2013dva,Detmold:2010ts, Engelhardt:2007ub,Christensen:2004ca}, HB$\chi$PT fit \cite{Griesshammer:2015ahu,Myers:2014ace,McGovern:2012ew}, B$\chi$PT fit \cite{Lensky:2014efa}, HB$\chi$PT \cite{Bernard:1993ry,Holstein:1999uu,Nevado:2007dd}, B$\chi$PT $\delta$-expansion \cite{Lensky:2015awa,Lensky:2014dda, Alarcon:2020wjg,Alarcon:2020icz} and B$\chi$PT $\epsilon$-expansion \cite{Bernard:2012hb,Thurmann:2020mog}.

Most recent HB$\chi$PT \cite{McGovern:2012ew,Griesshammer:2012we,Myers:2014ace,Myers:2015aba,Griesshammer:2015ahu} and B$\chi$PT \cite{Lensky:2008re,Lensky:2009uv,Lensky:2014efa, Lensky:2014dda,Lensky:2015awa,Lensky:2016nui,Alarcon:2020wjg, Alarcon:2020icz} calculations and fits of
CS observables employ the 
$\de$-expansion power counting. An exception are the works of Bernard et al.\ \cite{Bernard:2012hb} and Th\"urmann et al.\ \cite{Thurmann:2020mog}. As one can see from Figure~\ref{LTPolarizabilities} (upper panel), B$\chi$PT predictions for $\delta_{LTp}$ within the $\delta$-expansion \cite{Lensky:2014dda,Alarcon:2020icz} or the $\epsilon$-expansion \cite{Bernard:2012hb,Thurmann:2020mog} deviate substantially, since they include the $\Delta(1232)$ in different ways. In the $\epsilon$-expansion, the longitudinal-transverse polarizability receives a large contribution from diagrams where the photons couple directly to the $\Delta(1232)$ inside a loop. These diagrams are absent in the $\delta$-expansion at $\mathcal{O}(p
^4/\varDelta)$, thus, there the effect of the $\Delta(1232)$ is small and agrees with the MAID model \cite{MAID}. For the generalized longitudinal-transverse polarizability $\delta_{LTp}(Q^2)$ a similar $Q^2$ evolution is found in both power-counting schemes, see Figure \ref{Fig:deltaLTQ2} (left panel). Therefore, the discrepancy found for the polarizability  $\delta_{LTp}$ at the real-photon point continues as a constant shift for all $Q^2$ \cite{Alarcon:2020icz}. Another difference between the B$\chi$PT calculations
\cite{Lensky:2014dda,Alarcon:2020icz,Bernard:2012hb,Thurmann:2020mog} is the implementation of the magnetic-dipole $\mathcal{N}$-to-$\Delta$ transition and the coupling $g_M$ \cite{Krebs:2019ddp}. This ``$\delta_{LT}$ puzzle'' 
could soon be resolved by an empirical evaluation based on new data for the proton spin structure function $g_2$ from the Jefferson Lab ``Spin Physics Program''. A preliminary analysis \cite{Slifer:2018talk} favored the $\delta$-expansion power counting \cite{Alarcon:2020icz}, just like the MAID model does, cf.\ Figures \ref{LTPolarizabilities} and \ref{Fig:deltaLTQ2}. Note that the $\delta$-expansion results in Refs.~\cite{Alarcon:2020icz} and \cite{Lensky:2014dda} are both $\mathcal{O}(p^4/\varDelta)$. They differ by an improved error estimate and inclusion of the Coulomb coupling $g_C$ \cite{Alarcon:2020icz}. The $\epsilon$-expansion results in Refs.~\cite{Bernard:2012hb} and \cite{Thurmann:2020mog} are $\mathcal{O}(\epsilon^3)$ and $\mathcal{O}(\epsilon^3+p^4)$, respectively.

\begin{figure}[t]
\centering
\hspace{-1.5cm}\includegraphics[width=0.9\columnwidth]{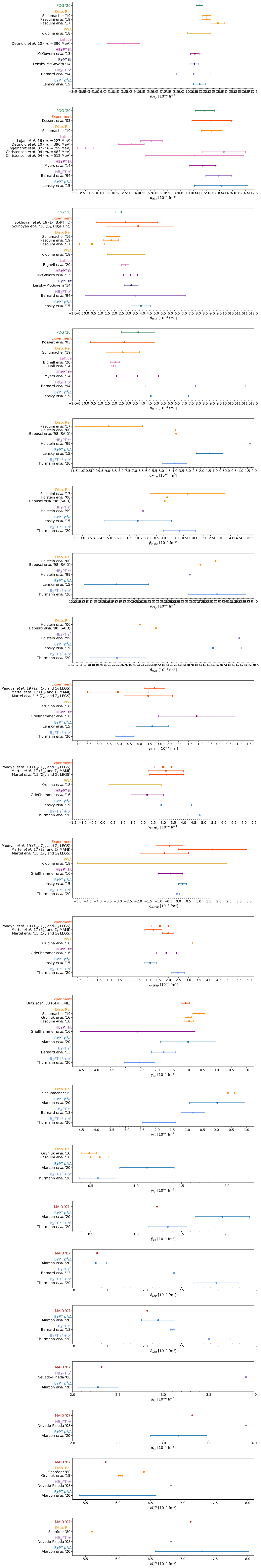}
\caption{Summary for the forward spin polarizability of the proton $\ga_{0p}$ (upper panel) and neutron $\ga_{0n}$ (lower panel). Theoretical predictions from chiral EFT are compared with empirical evaluations of the forward spin polarizability sum rule \eref{g0gen} at the real-photon point. \label{Gamma0Pol}}
\end{figure}

Similarly, we observe that the extensive set of empirical evaluations of the generalized forward spin polarizability $\ga_0(Q^2)$ at $Q^2<0.3$ GeV$^2$ agrees perfectly with the $\delta$-expansion prediction \cite{Alarcon:2020icz}, but differs from the $\epsilon$-expansion prediction \cite{Bernard:2012hb,Thurmann:2020mog}, cf.\ Figures \ref{Gamma0Pol} and \ref{Fig:gamma0plot} (upper panel). For the higher-order analogue $\bar \ga_0(Q^2)$, shown in Figure \ref{Fig:gamma0plot} (lower panel), the situation is less obvious. Only the dispersive evaluations of $\bar \ga_{0p}$ at the real-photon point, cf.\ Figure \ref{BarGamma0Pol}, are in slight disagreement with the $\mathcal{O}(p^4/\varDelta)$ prediction \cite{Alarcon:2020icz}, while conform with the $\mathcal{O}(\epsilon^3+p^4)$ prediction \cite{Thurmann:2020mog}.

The most studied polarizabilities are the electric and magnetic dipole polarizabilities, for which the Particle Data Group publishes recommended values \cite{Zyla:2020}. They are needed as input for calculations of the proton-structure effects  in the muonic-hydrogen Lamb shift from two-photon exchange. Of particular importance is $\beta_{M1p}$. It enters the $\ol T_1(0,Q^2)$ subtraction function \eref{subtractionfunction}, which has to be modeled \cite{Birse:2012eb} or predicted within $\chi$PT \cite{Alarcon:2020wjg,Lensky:2017bwi,Peset:2014jxa} because it cannot be measured in experiment or reconstructed from the unpolarized proton structure function $f_1$ in the dispersive approach, cf.\ \Eqref{T1dr}. Recently,  $\beta_{M1p}$ has therefore been extracted from the linear polarization beam asymmetry,
\beq
\label{Eq:sigma_3}
\Si_3 = \frac{ \dd\sigma_{||} - \dd\sigma_{\perp} }{ 
\dd\sigma_{||} + \dd\sigma_{\perp} },
\eeq
 measured for the proton by the A2 Collaboration \cite{Sokhoyan:2016yrc} and LEGS \cite{Blanpied:2001ae}. Up to $\mathcal{O}(\nu^2$), the beam asymmetry $\Si_3$ provides access to 
$\beta_{M1}$ independent of $\alpha_{E1}$
\cite{Pascalutsa13}:
\begin{flalign}\label{Eq:krupina}
\ol \Si_3 = - 
\frac{4 M_N \omega_B^2 \cos\theta_B \sin^2\theta_B }{
 (1+\cos^2\theta_B)^2 }\, \alpha^{-1}\beta_{M1}.
\end{flalign}
Presently, the extraction of $\beta_{M1p}$ from $\Sigma_3$  \cite{Sokhoyan:2016yrc} is not competitive with the standard dispersive analyses of unpolarized CS cross sections.
New high-precision measurements with significantly higher statistics should change this.

\begin{figure}[t]
\centering
\hspace{-1.9cm}\includegraphics[viewport=0 0 1115 475,clip,width=0.9\columnwidth]{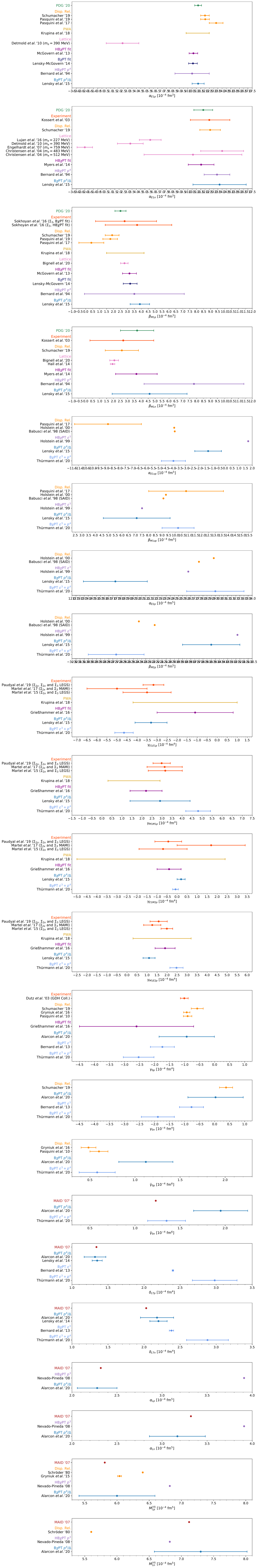}
\caption{Summary for the fourth-order Baldin sum rule of the proton $M_{1p}^{(4)}$ (upper panel) and neutron  $M_{1n}^{(4)}$ (lower panel). Theoretical predictions from chiral EFT are compared with empirical evaluations of the fourth-order Baldin sum rule \eref{SumRuleBaldin4} at the real-photon point. \label{Baldin4Pol}}
\end{figure}

Analyses of CS data with fixed-$t$ unsubtracted dispersion relations can be found in Refs.~\cite{Schumacher:2005an,Babusci:1998ww}, with an update in Ref.~\cite{Schumacher:2019ikn}. Fixed-$t$ subtracted dispersion relations are used in Ref.~\cite{Holstein:1999uu}, and are applied together with a bootstrap-based fitting technique in the recent Ref.~\cite{Pasquini:2019nnx}. Unfortunately, the dispersive and $\chi$PT fits tend to disagree for certain polarizabilities, e.g., for  $\al_{E1p}$ and $\beta_{M1p}$, cf.\ Figures~\ref{alphaE1Pol} and \ref{betaM1Pol} (upper panels). The $\mathcal{O}(p^4/\Delta)$ B$\chi$PT prediction \cite{Lensky:2015awa} and the B$\chi$PT fit \cite{Lensky:2014efa} of the proton dipole polarizabilities, see Figures~\ref{alphaE1Pol} and \ref{betaM1Pol} (upper panels), are in good agreement. A HB$\chi$PT fit, which also includes the lowest-order spin polarizabilities in Figures~\ref{ForwardSpinPolarizabilities} and \ref{Gamma0Pol}, agrees with the B$\chi$PT results \cite{Lensky:2015awa,Lensky:2014efa} except for $\ga_{M1E2p}$. Recently, a model-independent 
PWA of proton RCS data below pion-production threshold has shown \cite{Krupina:2017pgr} that the differences between dispersive approaches and B$\chi$PT results are due to inconsistent experimental data subsets, rather than the ``model-dependence'' of the theoretical frameworks. In the summary figures for the dipole and lowest-order spin polarizabilities, cf.\ Figures \ref{alphaE1Pol}, \ref{betaM1Pol} and \ref{ForwardSpinPolarizabilities} (upper panels), I show the spread of results from their PWA fits of different data subsets \cite{Krupina:2017pgr}. Note that all fits use the data-driven evaluations of the Baldin and forward spin polarizability sum rules from Refs.~\cite{Gryniuk:2016gnm, Gryniuk:2015eza} as input. Their analysis shows that the difference of proton scalar polarizabilities is constrained to a rather broad interval \cite{Krupina:2017pgr}: $\al_{E1p}- \beta_{M1p}=(6.9 \dots 10.9)\times 10
^{-4}$fm$^3$. In Ref.~\cite{Pasquini:2017ehj}, the dipole dynamical polarizabilities entering the multipole decomposition of the scattering amplitudes were for the first time extracted from proton RCS data below pion-production threshold. At lowest order, they are related to the static dipole and dispersive polarizabilities, see Figure~\ref{DispersivePol} (upper panel).

Both the partial-wave and the dispersive analysis in Refs.~\cite{Krupina:2017pgr} and \cite{Pasquini:2017ehj} come to the conclusion that quantity and quality of the CS data has to increase for improved extractions of the nucleon polarizabilities. A trend is going towards the measurement of beam asymmetries, such as $\Sigma_3$, and double-polarization observables:
\begin{subequations}
\beq
\Sigma_{2x} = {  \dd\sigma^R_{+x}  -   \dd\sigma^L_{+x}   \over 
  \dd\sigma^R_{+x}  +  \dd\sigma^L_{+x}  },
  \eeq
  \begin{figure}[h]
\begin{center}
\includegraphics[width=0.9\textwidth]{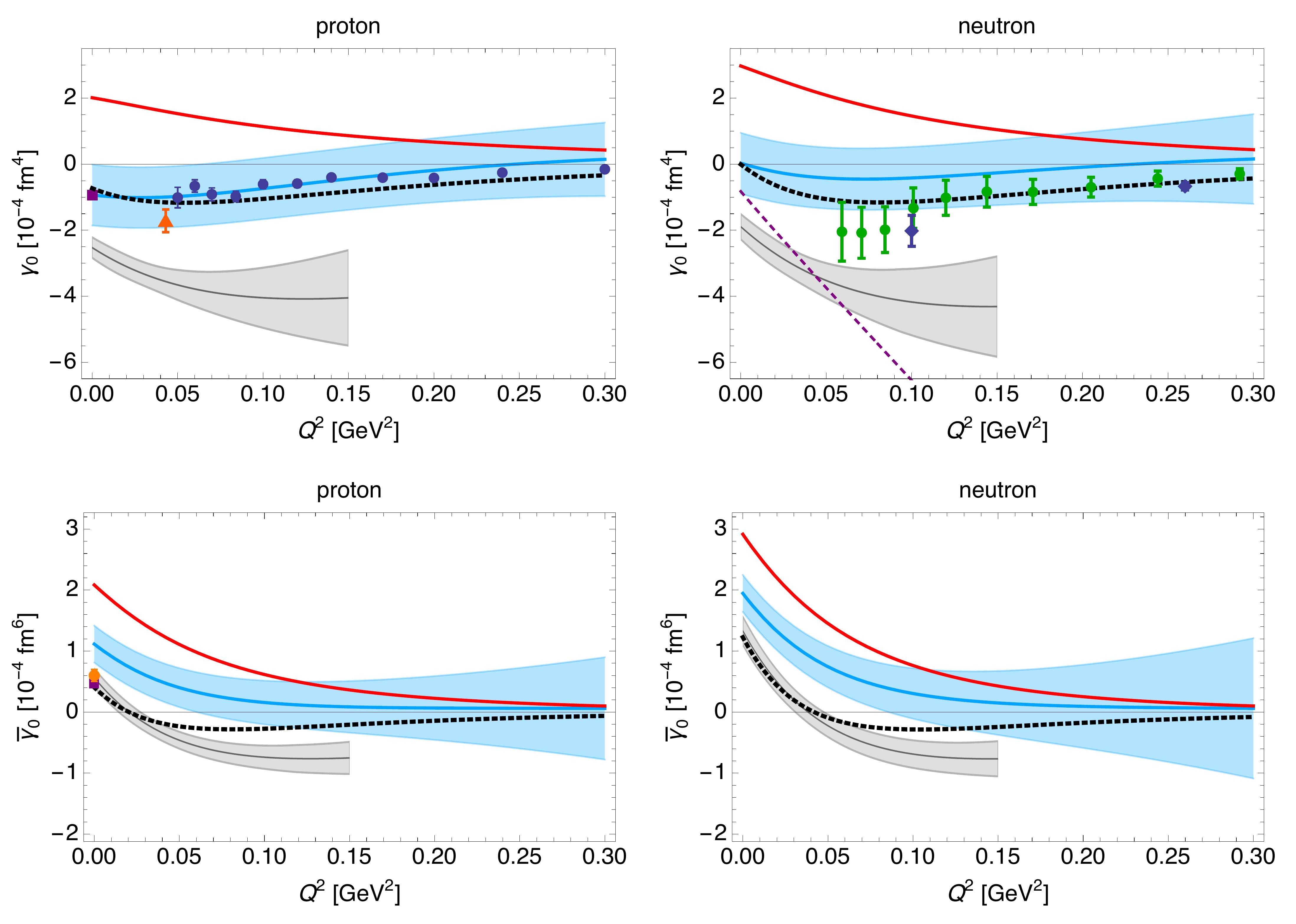}
\caption{Upper panel: Generalized forward spin polarizability, \Eqref{g0gen}, for the proton (left) and neutron (right) as function of $Q^2$. The black dotted line is the MAID model prediction~\cite{Drechsel:2000ct,Drechsel:1998hk,private-Lothar}, which is taken from Refs.~\cite{Drechsel:2002ar} (proton)
and~\cite{Amarian:2004yf} (neutron). The red line shows the leading-order B$\chi$PT result. The blue band is the $\mathcal{O}(p
^4/\Delta)$ B$\chi$PT result from  Ref.~\cite{Alarcon:2020icz}. The gray band is the $\mathcal{O}(\epsilon^3+p
^4)$ B$\chi$PT result from Ref.~\cite{Thurmann:2020mog}.
 The purple short-dashed lines is the $\mathcal{O}(p^4)$ HB results from Ref.~\cite{Kao:2002cp}; note that the corresponding proton curve is outside of the plotted range.
The experimental points for the proton are from: Ref.~\cite{Prok:2008ev} (blue dots), Ref.~\cite{Gryniuk:2016gnm} (purple square) and Ref.~\cite{Zielinski:2017gwp} (orange triangle; uncertainties added in quadrature). The experimental points for the neutron are from: Ref.~\cite{Amarian:2004yf} (blue diamonds) and Ref.~\cite{Guler:2015hsw} (green dots; statistical and systematic uncertainties added in quadrature). Lower Panel: Fifth-order generalized forward spin polarizability, \Eqref{barg0gen}, for the proton (left) and neutron (right) as function of $Q^2$. The black dotted line is the MAID model prediction \cite{MAID}. The experimental points for the proton are from: Ref.~\cite{Gryniuk:2016gnm} (purple square) and Ref.~\cite{Pasquini:2010zr} (orange dot).\label{Fig:gamma0plot}}
\end{center}
\end{figure}
  \beq
\Sigma_{2z} = {  \dd\sigma^R_{+z}  -   \dd\sigma^L_{+z}   \over 
  \dd\sigma^R_{+z}  +  \dd\sigma^L_{+z}  },
\eeq
\end{subequations}
where $\dd\sigma^{R(L)}_{+x}$ and $\dd\sigma^{R(L)}_{+z}$ are the differential cross sections for right (left) circularly polarized photons scattering from a nucleon target polarized either in the transverse $+\hat{x}$ direction or in the incident beam direction $+\hat{z}$. Here, the advantage is that systematic uncertainties, e.g., variations in photon flux or uncertainties in target thickness, are canceling out. Combining double-polarization observable and beam-asymmetry measurements, one is sensitive to the lowest-order spin polarizabilities, see Figure~\ref{ForwardSpinPolarizabilities}. For the extraction of the polarizabilities from the MAMI data for $\Sigma_{2x}$ \cite{Martel:2017pln,Martel:2014pba}, $\Sigma_{2z}$ \cite{Paudyal:2019mee} and $\Si_3$ \cite{Sokhoyan:2016yrc}, as well as the older LEGS data for $\Si_3$ \cite{Blanpied:2001ae}, one can use dispersive models \cite{Pasquini:2007hf,Holstein:1999uu,Drechsel:2002ar} or $\chi$PT fits \cite{Lensky:2009uv}.

\begin{figure}[t]
\centering
\hspace{-1.6cm}\includegraphics[viewport=0 500 1115 895,clip,width=0.9\columnwidth]{Figures/AllPolarizabilities05.pdf}
\caption{Summary for the longitudinal polarizability of the proton $\al_{Lp}$ (upper panel) and neutron $\al_{Ln}$ (lower panel). Theoretical predictions from chiral EFT are compared with the MAID unitary isobar model. \label{LongitudinalPolarizabilities}}
\end{figure}

Besides experimental efforts,
lattice QCD is making considerable progress. Most notably are the lattice QCD predictions for $\beta_{M1}$ with chiral extrapolation to physical pion mass \cite{Bignell:2018acn,Bignell:2020xkf}, as well as the plentiful results for $\al_{E1n}$ \cite{Lujan:2016ffj,Detmold:2010ts,Engelhardt:2007ub,Christensen:2004ca}. By now, even direct lattice evaluations of the unpolarized forward VVCS amplitudes became possible and can be used to determine the structure functions and their moments \cite{Hannaford-Gunn:2020pvu,Chambers:2017dov,Can:2020sxc}.

In Figures~\ref{LTPolarizabilities}, \ref{BarGamma0Pol}, \ref{Gamma0Pol}, \ref{Baldin4Pol}, \ref{LongitudinalPolarizabilities}, one can see updated results from the recent $\mathcal{O}(p^4/\Delta)$ B$\chi$PT prediction of unpolarized VVCS \cite{Alarcon:2020wjg}, related to $\al_L$ and $M_1^{(4)}$, and polarized VVCS \cite{Alarcon:2020icz}, related to $\delta_{LT}$, $\ga_0$ and $\bar \ga_0$. The latter could be compared to new results from the Jefferson Lab ``Spin Physics Program'' for the proton spin structure functions $g_1$ and $g_2$, see for instance the E08-027 experiment \cite{Zielinski:2017gwp} and the E97-110 experiment \cite{Sulkosky:2019zmn}. Note that the HB$\chi$PT predictions for $M_1
^{(4)}$ and $\al_L$ shown in Figures \ref{Baldin4Pol} and \ref{LongitudinalPolarizabilities} were extracted from the VVCS amplitudes presented in Ref.~\cite{Nevado:2007dd}, but are not quoted in the original work.

\section{Conclusion and Outlook} \label{hydrogen}

The chiral EFT expansion for nucleon polarizabilities begins with inverse powers of pion
mass and other light scales, such as the nucleon-$\Delta$ mass difference. 
These inverse powers ($1/m_\pi$, $1/\varDelta$, etc.) along with the chiral logs  
constitute predictions of $\chi$PT. As such, the polarizabilities, and, in fact, the entire
process of CS at low energies, provide a testing ground for $\chi$PT.

Moreover, the interpretation of low-energy CS data and the extraction of  nucleon polarizabilities therefrom should rely on a systematic theoretical framework such as $\chi$PT. In what we have seen thus far, $\chi$PT is quite successful in the
prediction of nucleon polarizabilities. It can as well be used to design ``optimal'' future experiments  for improving the empirical determinations of nucleon polarizabilities \cite{Melendez:2020ikd}. 

An alternative to $\chi$PT, in the field of nucleon CS, is provided by models based on fixed-$t$ dispersion relations~\cite{Lvov:1996xd,Drechsel:1999rf}. The theoretical uncertainties of the dispersive approach are harder to understand,  but, at least within the quoted uncertainties, the extracted values of polarizabilities are in overall comparable to those
found in $\chi$PT. However, a few discrepancies remain. 
For example, the tension in the value of the proton magnetic dipole polarizability still persists, cf.\ ``Disp.~Rel.'' vs.\ $\chi$PT results in Figure~\ref{betaM1Pol} (upper panel). 
A model-independent PWA shows~\cite{Krupina:2017pgr} that this discrepancy is likely to be caused by the experimental CS database, rather than the differences between the theoretical frameworks. With MAMI  \cite{Mornacchi:2019rtq}  and HIGS~\cite{Ahmed:2020hux} experiments underway, the database will soon be greatly improved. It is worth mentioning that MAMI is also finalising a program to measure the CS double-polarization observables ($\Sigma_{2x}$, $\Sigma_{2z}$) which will lead to an improved
extraction of proton spin polarizabilities \cite{Martel:2017pln,Martel:2014pba,Paudyal:2019mee}.

Even among the various $\chi$PT calculations there are significant discrepancies 
that need to be understood. 
The differences between the heavy-baryon (HB$\chi$PT) and the Lorentz-invariant covariant (B$\chi$PT) results
are not difficult to track. However, differences among various B$\chi$PT calculations are more troublesome. A prominent example is the longitudinal-transverse polarizability of the proton (upper panel of Figure~\ref{LTPolarizabilities} and left panel of Figure \ref{Fig:deltaLTQ2}), where the $\delta$- and $\epsilon$-expansion  B$\chi$PT calculations are different by about a factor of 2. This ``$\delta_{LT}$ puzzle'' 
could soon receive an experimental resolution, when the long-promised data from Jefferson Lab ``Spin Physics Program'' \cite{Zielinski:2017gwp,Adhikari:2017wox,Sulkosky:2019zmn}  on the proton spin structure function $g_2$ will be published \cite{Slifer:2018talk}. Besides the polarizabilities, the Gerasimov--Drell--Hearn sum rule for the neutron will be verified by the E97-110 experiment using a helium-3 target \cite{Ton:2019rci}.

In the mean time, lattice QCD calculations of nucleon polarizabilities are advancing towards the
physical pion mass. Until now, however,  $\chi$PT has been used to extrapolate the lattice results to the physical mass \cite{Bignell:2020xkf,Hall:2013dva}. A significant progress has recently been achieved in calculating the proton polarizabilities \cite{Bignell:2020xkf,Detmold:2010ts}, and in direct calculations of the spin-independent forward VVCS \cite{Hannaford-Gunn:2020pvu,Chambers:2017dov,Can:2020sxc}.

In the next few years, one can expect a lot of progress in this field, mainly
due to the upcoming data from MAMI, HIGS and Jefferson Lab. New 
$\chi$PT and lattice QCD calculations will certainly continue to advance and will, hopefully, 
bring some clarity on the aforementioned discrepancies.

\vspace{6pt} 

\funding{Financial support from the Swiss National Science Foundation is gratefully acknowledged.}

\acknowledgments{I would like to thank Jose~M.~Alarc{\'o}n, Vadim~Lensky, Vladimir~Pascalutsa and Marc~Vanderhaeghen for the fruitful collaboration on this topic, and Gilberto~Colangelo for many useful remarks on the manuscript. }

\abbreviations{The following abbreviations are used in this manuscript:\\

\noindent 
\begin{tabular}{@{}ll}
B$\chi$PT & Baryon chiral perturbation theory\\
$\chi$PT & Chiral perturbation theory\\
CS & Compton scattering\\
EFT & Effective-field theory\\
HB$\chi$PT & Heavy-baryon chiral perturbation theory\\
LEC & Low-energy constant\\
PWA & Partial-wave analysis\\
RCS & Real Compton scattering\\
VCS & Virtual Compton scattering\\
VVCS & Forward doubly-virtual Compton scattering\\
\end{tabular}}

\reftitle{References}
\bibliography{lowQ}

\end{document}